\pdfoutput=1
\documentclass[10pt,twocolumn,a4paper]{article}
\usepackage[T1]{fontenc}
\usepackage{lmodern}
\usepackage{amsmath, amssymb, amsfonts}
\usepackage{graphicx}
\usepackage{cite}
\usepackage{booktabs}
\usepackage{geometry}
\usepackage{titlesec}
\usepackage{caption}
\usepackage{float}
\usepackage{bm}
\usepackage{tikz}
\usepackage[american]{circuitikz}
\usetikzlibrary{arrows.meta, positioning, calc, shadows, shapes.geometric}
\usepackage{pgfplots}
\usepgfplotslibrary{fillbetween}
\pgfplotsset{compat=1.15}
\usetikzlibrary{intersections, fillbetween, patterns}

\titleformat{\section}{\large\bfseries}{\arabic{section}.}{1em}{}
\titleformat{\subsection}{\normalsize\bfseries}{\arabic{section}.\arabic{subsection}}{1em}{}
\titleformat{\subsubsection}{\small\bfseries}{\arabic{section}.\arabic{subsection}.\arabic{subsubsection}}{1em}{}

\title{\Large \textbf{Rapid Heat Removal from Chip Cores: Near-Junction Active Cooling of Power Transistors via Electron Entropy Transport and Spatial Decoupling}}
\author{Xiangying Li$^{1}$\\[1ex] \textit{Independent Researcher} (Guangzhou 510000)}
\date{}

\begin{document}

\maketitle

\begin{abstract}
How do you pull heat out of a chip's core fast enough to hold the temperature line? This is a shared bottleneck for large-scale integrated circuits and high power-density devices. We study the hardest case: RF power GaN HEMTs, whose local hotspots exceed $5\,\mathrm{kW/cm^2}$ in heat flux. A horizontal near-junction active cooling topology is proposed and analyzed, built on electron entropy transport. Classical on-chip thermoelectric cooling suffers from a contest at the interface. Peltier absorption grows with current ($\propto I$), while Joule heating grows with its square ($\propto I^2$). In a vertical thermoelectric pillar, the pillar's own conduction resistance locks about half of the Joule heat back toward the cold end. That sets a rigid ceiling on net cooling. Working from the Onsager framework of nonequilibrium thermodynamics, we show where the asymmetry comes from. Peltier absorption is an interface source: the term $T\bm{J}\cdot\nabla\alpha$ localizes where the Seebeck coefficient jumps. Joule heating ($\rho J^2$) fills the volume. The topology exploits this fact. A high-performance thermoelectric thin film lies flat on a highly conductive diamond substrate, right beside the drain (the electron exit) of the device. Bulk Joule heat then leaves along the shortest vertical path, into the substrate. We solve a lateral fin model for the film and derive a closed form for the recirculation factor $\gamma$. The vertical pillar gives $\gamma \approx 1$. The horizontal film gives $\gamma_{\max} \approx \sqrt{2}\,L_z/L_x$. With typical values ($L_z = 10\,\mathrm{\mu m}$, $L_x = 500\,\mathrm{\mu m}$), $\gamma \approx 0.03$. The decoupling scales both the net cooling ceiling and the critical runaway current $\lambda_m$ by roughly $1/\gamma$. At equal resistance, the same film lifts maximum cooling from about $13\,\mathrm{mW}$ to about $500\,\mathrm{mW}$, and $\lambda_m$ from $0.65\,\mathrm{A}$ to $24\,\mathrm{A}$. A time-constant analysis (electron response $\sim\mathrm{ns}$, film recirculation $\sim 40\,\mathrm{\mu s}$, substrate $\sim 0.1\,\mathrm{s}$) and a linear stability analysis with temperature-sensitive parameters define the working window and safety margins for microsecond pulse cooling. Electrical isolation between the film and the drain, with its parasitics, is also discussed. The architecture leaves the intrinsic semiconductor structure untouched. The principle, absorb heat at an interface and extract bulk heat vertically, does not depend on a particular device family. It applies to any chip with a hot core, and offers a tractable, reproducible design path for near-junction thermal management.

\noindent \textbf{Keywords:} chip thermal management; large-scale integrated circuits; power transistors; GaN HEMT; near-junction thermal transport (NJTT); Peltier effect; electron entropy transport; spatial decoupling; recirculation factor
\end{abstract}

\section{Introduction}
From high-compute processors to power transistors, chip performance is increasingly limited by the same physical bottleneck: the core region produces heat faster than it can be moved out. Wide-bandgap semiconductors such as GaN and SiC are now in wide use. Modern power converters, RF amplifiers, and large-scale integrated circuits push toward ever higher frequencies and power densities. Devices shrink while integration grows, so microscopic heat flux concentrates inside the chip. Under extreme RF power or transient pulse conditions, severe local hotspots form near the 2DEG channel of a GaN HEMT. The peak heat flux at these microscopic cores reaches $5\,\mathrm{kW/cm^2}$, more than an order of magnitude above the chip-average flux\cite{khanna2017thin}.

Traditional remedies (heat pipes, micro-channel liquid cooling, phase-change materials) are macroscopic and passive. Heat must pass in sequence through the semiconductor substrate, the thermal interface material (TIM), the vapor chamber, and the heat sink. Such a conduction chain has two built-in flaws. First, many dissimilar interfaces stack up thermal resistance. Second, conduction carries a thermal time constant; it is slow. As a result, external cooling cannot reach down to the micron-scale junction in time. Even with the liquid loop running at full load, the internal hotspot stays hot. Carrier mobility drops, dynamic on-resistance ($R_{ds(on)}$) degrades, and in the worst case the device suffers irreversible electro-thermal breakdown or thermal runaway. Large-scale integrated circuits fail the same way: throttling, rising leakage, and accelerated aging all trace back to loss of control over core temperature. For this reason, near-junction thermal transport (NJTT), active cooling applied at the heat source itself and bypassing the package stack, has become a frontier direction for breaking the thermal bottleneck of power devices\cite{bar2012darpa}.

On-chip thermoelectric cooling (TEC) has drawn wide attention in recent years. But micro-scale TEC runs into a basic physical constraint as it approaches high-power hotspots. Peltier absorption at the interface grows linearly with drive current ($\propto i$), while the accompanying Joule heating grows with the square ($\propto i^2$). At the large currents needed for $5\,\mathrm{kW/cm^2}$-class flux, Joule heat quickly dominates. In the classical vertical pillar, about half of this volumetric heat flows back to the cold end along the axial temperature gradient. Net cooling power then has a rigid ceiling set by both material and geometry\cite{long2010optimization}. How to press down that recirculation fraction, at the level of heat-flow topology, is the core physics question standing between micro-scale active cooling and engineering use.

This paper revisits the carrier behavior in micro-scale electro-thermal coupling, starting from nonequilibrium thermodynamics and solid-state physics. At the microscopic level, current is the macroscopic appearance of directed charge motion. The entities that actually carry heat and energy and exchange energy with lattice phonons are electrons. On this basis, we propose a device-level intrinsic cooling topology\cite{li2011patent}. Its core idea: place the absorption interface near the ``electron exit'' of the device, that is, the drain.

More importantly, we analyze and quantify the ``spatial asymmetry'' of the two thermal effects in three dimensions. Within the Onsager transport framework, Peltier absorption corresponds to an interface-localized source in the energy equation: $T\bm{J}\cdot\nabla\alpha$ degenerates to a $\delta$ function at a material interface. Joule heating ($\rho J^2$), by contrast, is a volume-distributed source. Building on this, we lay the thermoelectric film flat on a high-conductivity diamond substrate, and set up a lateral fin model for the arrangement. An analytic expression for the Joule recirculation factor $\gamma$ follows. The vertical pillar locks $\gamma$ at $1/2$ through the internal conduction resistance (corresponding to $\gamma\approx 1$ under the convention of this paper). The horizontal film gives a $\gamma_{\max}$ set only by the thickness-to-length ratio $L_z/L_x$, down to the $10^{-2}$ range. We further derive the generalized critical current $\lambda_m$ and the maximum temperature difference $\Delta T_{max}(\gamma)$. A linear stability analysis with temperature-sensitive parameters (TSP) gives the dynamic runaway boundary. A time-constant analysis defines the window for microsecond pulse cooling. These criteria rest only on where the heat sources sit in space, not on any particular materials system. The design logic transfers to any chip with a hot core, and provides an analytic, reproducible set of physical rules for self-cooled device design.

\section{Related Work and Technology Evolution}
For local hotspots in high power-density ICs and wide-bandgap devices, existing work spans near-junction thermal transport, diamond heterointegration, thin-film thermoelectric cooling, and electro-thermal modeling. This section reviews the threads, to fix the position and the increment of the present architecture in ``spatial decoupling''.

\subsection{Near-Junction Thermal Transport (NJTT) and the Limits of Macroscopic Cooling}
As GaN HEMT power density rises, the thermal bottleneck moves from the package to the micron-scale region near the junction. Al-Areqi et al. reviewed thermal management for high power-density GaN converters and noted that, because of the very small die size, GaN devices carry far higher heat flux than SiC; traditional paths cannot get past the substrate conduction limit\cite{al2026review}. To reach the junction, Liu et al. embedded micro-channels inside a GaN-on-SiC substrate, an active near-junction scheme that confirmed the decisive role of physical proximity to the heat source, in saturation current and self-heating suppression\cite{liu2024near}. Such micro-fluidic approaches, however, carry high micro-fabrication cost and pumping loss, and are difficult to deploy at scale in ultra-high-frequency RF chips\cite{li2022near}.

\subsection{Diamond Heterointegration and the Thermal Boundary Resistance (TBR) Problem}
Diamond conducts heat better than any natural material ($>2000\,\mathrm{W/m\cdot K}$) and is an ideal passive substrate. Lu et al. argued for the large potential of integrating CVD diamond directly beneath GaN HEMTs\cite{lu2022advancement}. But Anaya et al. showed with physical probes that polycrystalline diamond carries severe nano-structural defects in the nucleation region. Phonon scattering intensifies at these interfaces and produces a high thermal boundary resistance (TBR). This thermal bottleneck erodes diamond's theoretical advantage\cite{anaya2019thermal}. To avoid the lattice stress of high-temperature bonding, Cheng et al. and Hardy et al. developed room-temperature van der Waals hetero-bonding and micro-transfer printing respectively, aiming at reliable GaN-diamond integration\cite{cheng2020interfacial, hardy2025micro}. The horizontal topology proposed here takes a different route. It uses a high-performance thermoelectric material as a ``heat pump'' first, then couples to diamond. Interface pumping combined with vertical extraction reduces reliance on a single phonon conduction path.

\subsection{Thin-Film Thermoelectric Cooling (TFTEC) and Transient Pulsing}
On-chip TEC based on the Peltier effect has advanced considerably. Benoit et al. used TEC to suppress local hotspots in coordination with a global cooling array, lowering overall energy\cite{benoit2016tecfan}. At the device level, Wu et al. established, through multi-physics simulation and experiment, the cooling performance boundary of multi-stage thin-film thermoelectric devices under idealized vacuum and no-plate conditions\cite{wu2024validation}. At the materials level, superlattice thin-film modules use nanoscale periodicity to achieve low thermal conductivity with high electrical conductivity, raising the figure of merit $ZT$. The DARPA-led ICECool program verified the engineering feasibility of growing $\mathrm{Bi_2Te_3}$ superlattice films, $5\sim 20\,\mathrm{\mu m}$ thick, directly near the junction of high-power GaN amplifiers\cite{khanna2017thin, bulman2016superlattice}. On the transient side, Manno and Yang's groups studied ``transient pulse cooling'' as a nonequilibrium process. They showed that short current pulses well above steady-state limits produce interfacial supercooling and markedly cut peak hotspot temperature\cite{manno2015transient, yang2022transient}. This paper combines that time-scale difference with a spatially decoupled architecture, and gives a quantitative working window that keeps Joule recirculation from taking over.

\subsection{Classical Electro-Thermal Models and the Joule Recirculation Paradox}
The physical core of TEC is the Peltier effect. When carriers cross a heterojunction between different materials, the difference in Seebeck coefficient (and Fermi level) makes them exchange energy with the lattice, absorbing or releasing heat at the interface\cite{long2010optimization}. The classical one-dimensional compact model gives the cold-end heat flux $q_c$ as three terms:
\begin{equation}
q_c = \alpha i \theta_c - \frac{1}{2} r i^2 - \kappa(\theta_h - \theta_c)
\label{eq:classic}
\end{equation}
Here $\alpha$ is the Seebeck coefficient, $r$ the internal resistance, $\kappa$ the thermal conductance, $i$ the drive current, and $\theta_c$, $\theta_h$ the cold- and hot-end temperatures. The middle term, $-\frac{1}{2} r i^2$, exposes the structural weakness of the vertical TEC: of the Joule heat generated inside the pillar, about half conducts back to the cold end, because the resistance distribution is symmetric. Section \ref{sec:gamma} proves that this half-split is not an empirical assumption, but a strict consequence of the pillar's internal conduction resistance. At micron scale and high flux, the very short axial distance makes this return term especially damaging to net cooling.

\subsection{Electro-Thermal Coupling in Wide-Bandgap Devices and Dynamic Search}
Under extreme electro-thermal coupling, physical parameters drift. Rahman et al. built a mechanism-oriented degradation model and analyzed how temperature-sensitive parameters (TSEPs) affect charge trapping dynamics and dynamic on-resistance in wide-bandgap power transistors\cite{rahman2026mechanism}. On the control side, Jieyi Long et al. formulated a convex program based on inverse-matrix theory and proved the existence of a critical current ceiling $\lambda_m$ in TEC systems: beyond it, the system diverges thermodynamically\cite{long2010optimization}. Existing greedy deployment algorithms, however, stay within the vertical micro-bump geometry. They never separate the $\propto i$ and $\propto i^2$ heat sources into different spatial dimensions. That is exactly the problem we address.

\section{Nonequilibrium Thermodynamics of Electron Entropy Transport}
This section builds the theoretical basis of the paper. We give a strict definition of ``electron entropy transport'', trace its origin in the spatial asymmetry between interface-localized Peltier absorption and volume-distributed Joule heating, and recall the classical limits of thermoelectric cooling as the baseline for comparison.

\subsection{The Onsager Framework and a Strict Definition of Electron Entropy Flow}
Nonequilibrium thermodynamics describes the coupling of current and heat flow as a linear relation between generalized fluxes and forces through an Onsager transport matrix. For an isotropic thermoelectric medium, the charge current density $\bm{J}_e$ and heat current density $\bm{J}_q$ satisfy:
\begin{equation}
\begin{pmatrix} \bm{J}_e \\[2pt] \bm{J}_q \end{pmatrix}
=
\begin{pmatrix} \sigma & \sigma\alpha \\[2pt] \sigma\alpha T & \sigma\alpha^2 T + \kappa_e \end{pmatrix}
\begin{pmatrix} \bm{E} \\[2pt] -\nabla T \end{pmatrix}
\label{eq:onsager}
\end{equation}
where $\sigma$ is electrical conductivity, $\alpha$ the Seebeck coefficient, $\kappa_e$ the electronic thermal conductivity, and $\bm{E}$ the electric field. The Onsager reciprocal relation requires the off-diagonal terms to satisfy $\sigma\alpha T = T\cdot(\sigma\alpha)$, which yields the \textbf{Kelvin relation} between Peltier and Seebeck coefficients:
\begin{equation}
\Pi = \alpha T
\label{eq:kelvin}
\end{equation}
From the convective part of the heat flux, the entropy current carried by the charge flow follows directly:
\begin{equation}
\bm{J}_s^{\,e} = \frac{\bm{J}_q^{\,\mathrm{conv}}}{T} = \alpha\,\bm{J}_e
\label{eq:entropy}
\end{equation}
Equation (\ref{eq:entropy}) states the precise meaning of ``electron entropy transport'': \textbf{the Seebeck coefficient is the entropy transported per unit charge}, an entropy-to-charge ratio. A current carries charge, and it also carries entropy; $\alpha$ measures how much. Microscopically, the Mott relation ties $\alpha$ to the mean carrier entropy within the transport window, so a good thermoelectric material is one whose carriers carry more entropy per unit charge. When electrons flow from a low-$\alpha$ material (metal, $|\alpha_m|\sim \mathrm{\mu V/K}$) into a high-$|\alpha|$ material ($\mathrm{Bi_2Te_3}$, $|\alpha|\sim 200\,\mathrm{\mu V/K}$), the entropy stream carried by the electron flow must jump. The lattice phonon system supplies the difference. Macroscopically, the interface absorbs heat.

\subsection{The Spatial Asymmetry of the Two Heat Sources: A Strict Argument}
Substituting the constitutive relations (\ref{eq:onsager}) into the steady energy balance $\nabla\cdot\bm{J}_q = \bm{J}_e\cdot\bm{E}$ gives the heat equation with current:
\begin{equation}
\nabla\cdot(k\nabla T) + \rho \bm{J}^2 - T\,\bm{J}\cdot\nabla\alpha = 0
\label{eq:source}
\end{equation}
Equation \eqref{eq:source} exposes the essential asymmetry in the spatial character of the two heat sources:
\begin{enumerate}
\item \textbf{The Joule term $\rho J^2$}: it depends on the square of current density and fills the whole volume the current traverses. It is a \textbf{volume-distributed heat source}.
\item \textbf{The Thomson--Peltier term $T\bm{J}\cdot\nabla\alpha$}: it appears only where the Seebeck coefficient varies in space. At a hetero-interface, $\alpha$ steps; $\nabla\alpha$ degenerates to a $\delta$ function at the interface. Integrating along the interface normal gives the areal absorption rate:
\begin{equation}
Q_{peltier} = (\alpha_{TE} - \alpha_{m})\,T_c\, I
\label{eq:peltier}
\end{equation}
(The sign follows the current direction and carrier type; we take absorption as positive. For a metal--N-type thermoelectric junction, a proper current direction makes $\alpha_{TE}-\alpha_m$ correspond to absorption.)
\end{enumerate}
So the statement ``Peltier is an interface effect, Joule is a volume effect'' is not folklore. It is a direct consequence of the mathematical nature of the two source terms in the energy equation \eqref{eq:source}. One point deserves emphasis: the \textbf{generation} of Peltier absorption is strictly interfacial. But whether the pumped heat persists, and where the Joule heat goes, is decided by the topology of the whole thermal network. That is exactly the problem the spatial decoupling design in Section \ref{sec:topology} addresses.

\subsection{The Classical Limit of Thermoelectric Cooling (Baseline)}
In the classical vertical configuration, about half of the Joule heat returns to the cold end (Section \ref{sec:gamma} gives the proof). The net cooling power at the cold side is:
\begin{equation}
Q_{net} = \alpha T_c I - \frac{1}{2} I^2 R - K\Delta T
\label{eq:net-classic}
\end{equation}
where $R$ is the leg resistance, $K$ the leg thermal conductance, and $\Delta T = T_h - T_c$. Setting $\partial Q_{net}/\partial I = 0$ gives the classical limits:
\begin{equation}
I_{opt} = \frac{\alpha T_c}{R},\qquad
Q_{max} = \frac{(\alpha T_c)^2}{2R} - K\Delta T
\label{eq:limits}
\end{equation}
Setting $Q_{net} = 0$ gives the maximum temperature difference:
\begin{equation}
\Delta T_{max} = \frac{(\alpha T_c)^2}{2RK} = \frac{1}{2} Z T_c^2,\qquad
Z = \frac{\alpha^2}{RK}
\label{eq:dtmax-classic}
\end{equation}
The coefficient of performance is $\mathrm{COP} = Q_{net}/(\alpha I \Delta T + I^2 R)$. The classical result says: no matter how good the material ($ZT$), the half-split of Joule heat, a \textbf{topological constraint}, caps $Q_{max}$ and $\Delta T_{max}$ at the values in Eqs. \eqref{eq:limits}--\eqref{eq:dtmax-classic}. The central claim of this paper is that this cap is not a necessity of the second law. It is a consequence of one particular geometry, and a change of geometry can press it down.

\section{Horizontal Near-Junction Topology and the Spatial Decoupling Mechanism}
\label{sec:topology}
Based on the framework above, this section develops the device-level design of the horizontal decoupled topology, including the electrical interface, and derives the Joule recirculation factor $\gamma$ analytically.

\subsection{Topology Layout: Interface Pumping at the Electron Exit}
In a working GaN HEMT, electrons injected from the source are accelerated in the 2DEG channel by the gate field. When the fast electrons reach the drain (the electron exit), the channel field drops abruptly and the bands bend sharply. Hot electrons scatter strongly against optical phonons. This is the microscopic origin of extreme hotspots in high-frequency devices.

The core design of this work: near the drain metal contact, where heating is worst, integrate a thin film of high-Seebeck thermoelectric material (such as N-type $\mathrm{Bi_2Te_3}$ superlattice). By Eq. \eqref{eq:peltier}, when current crosses the metal electrode--thermoelectric interface, an areal absorption $(\alpha_{TE}-\alpha_m)T_c I$ appears there. The energy-level transition happens within a few lattice constants, so this Peltier absorption is strictly two-dimensional: a microscopic pumping interface hugging the heat source. Consistent with the framework of Section 3, we stress again: Peltier absorption is \textbf{generated} at the interface, but the performance of the whole cooling system depends on \textbf{where} the Joule volume heat and the returning heat flows end up. That is the object of the topology below.

\subsection{Electrical Interface and Independent Supply}
\label{sec:interface}
The electrical interface between the thermoelectric film and the drain is key to a device-level implementation. We adopt a reference design that is \textbf{electrically isolated but thermally coupled}:
\begin{enumerate}
\item \textbf{Vertical stack}: GaN epitaxy / drain metal (M1) --- ultra-thin insulating thermal bridge ($50\sim100\,\mathrm{nm}$ AlN or SiN) --- cold-side electrode of the film (M0) --- thermoelectric film (lying flat) --- diamond substrate. The areal thermal resistance of the bridge is about $d/\kappa_{AlN} \approx 10^{-7}/200 = 5\times10^{-10}\,\mathrm{m^2K/W}$, far below the interface TBR ($\sim10^{-8}\,\mathrm{m^2K/W}$). Thermally it is nearly transparent; electrically it provides complete isolation.
\item \textbf{Independent current loop}: the cooling current flows ``cold electrode M0 $\to$ laterally through the film body $\to$ far-side hot electrode $\to$ wiring on the diamond side $\to$ an independent driver''. It shares no path with the drain RF/power loop. Peltier absorption occurs at the M0/film interface, directly under the drain hotspot region.
\item \textbf{Parasitics}: the drain current path is untouched, so $R_{ds(on)}$ and series inductance are unaffected. The main parasitic is the overlap capacitance between the cold electrode and the drain metal, $C_{ov} \approx \varepsilon_0\varepsilon_r A_{ov}/d$. For full-pad coverage ($100\times100\,\mathrm{\mu m^2}$, $d=100\,\mathrm{nm}$, $\varepsilon_r\approx8$), $C_{ov}\approx 7\,\mathrm{pF}$ --- negligible for power switching (MHz). For RF ($10\,\mathrm{GHz}$), limit the overlap to a strip near the hotspot (down to $\sim0.5\,\mathrm{pF}$) or use a conformal design. In fact, the covering electrode can double as a gate field plate, turning a parasitic into an electrical benefit.
\end{enumerate}
For comparison, a direct-contact variant (drain metal metallurgically bonded to the film, with the cooling current crossing that interface) saves the insulator. But drain current then diverts into the film. The film sheet resistance $\rho/L_z \approx 1\,\Omega/\square$ can be comparable to the channel resistance, so the current division must be evaluated with care, and RF matching becomes harder. All analysis below uses the isolated reference design.

\subsection{Analytic Derivation of the Joule Recirculation Factor}
\label{sec:gamma}
This subsection is the theoretical core of the paper. We first prove that the vertical configuration's return fraction is locked by the internal resistance. We then set up a lateral fin model for the horizontal film and solve for the recirculation factor.

\subsubsection{The Vertical Pillar: A Half-Split Locked by Internal Resistance}
For a conventional vertical thermoelectric leg (height $L$, cross-section $A$), the volumetric Joule heating satisfies the one-dimensional steady Poisson equation:
\begin{equation}
k \frac{d^2T}{dz^2} + \rho J^2 = 0
\label{eq:slab}
\end{equation}
The solution is a parabolic temperature profile. The Joule-heat flux component at the cold end ($z=0$) is $-\rho J^2 L/2$: \textbf{exactly half} of the Joule heat flows toward the cold side. This holds strictly for any end temperatures (Dirichlet conditions). More generally, if the cold and hot ends connect to external sinks through areal resistances $R_c''$ and $R_h''$, the exact solution gives the fraction of Joule heat flowing to the cold end as:
\begin{equation}
\phi_{pillar} = \frac{R_h'' + \dfrac{L}{2k}}{R_c'' + R_h'' + \dfrac{L}{k}}
\label{eq:pillar-split}
\end{equation}
Equation \eqref{eq:pillar-split} has a transparent structure: the distributed volume source acts as if concentrated at the mid-plane of the pillar, and heat exits each way through half the leg resistance plus the end resistance. The order-of-magnitude point is what matters. For a film leg with $L = 10\,\mathrm{\mu m}$ and $k = 1.5\,\mathrm{W/mK}$, $L/2k \approx 3.3\times10^{-6}\,\mathrm{m^2K/W}$, two or more orders above the interface resistance achievable by advanced bonding ($\sim10^{-8}\,\mathrm{m^2K/W}$). Hence $\phi_{pillar} \approx 1/2$ is \textbf{locked by the pillar's own internal conduction resistance and barely responds to end conditions}. One conclusion follows: Joule recirculation in vertical TEC cannot be engineered away at interfaces. It must be solved by re-shaping the topology.

\subsubsection{The Horizontal Film: Lateral Fin Model and Closed-Form Recirculation Factor}
Our configuration lays the thermoelectric film flat, large-area, on high-conductivity diamond. The cold electrode sits at $x=0$; the film extends laterally to $L_x$, where the far end carries the hot electrode. Joule heat generates uniformly in the film (volume density $q''' = \rho J^2$) and has two ways out: straight down through the areal resistance $R_{th,z}''$ (half-thickness resistance $L_z/2k_z$, interface TBR, and diamond spreading resistance) into the diamond; or sideways to the cold electrode, which is recirculation. Let the overtemperature be $\theta(x) = T(x) - T_{dia}$. The lateral one-dimensional fin equation (per unit width) reads:
\begin{equation}
k_{xy} L_z \frac{d^2\theta}{dx^2} - \frac{\theta}{R_{th,z}''} + q''' L_z = 0
\label{eq:fin}
\end{equation}
where $k_{xy}$ is the in-plane thermal conductivity. Define the lateral thermal healing length and the Joule plateau overtemperature:
\begin{equation}
m = \frac{1}{l_p} = \frac{1}{\sqrt{k_{xy} L_z R_{th,z}''}},\qquad
\theta_p = q''' L_z R_{th,z}''
\label{eq:m-theta}
\end{equation}
$\theta_p$ has a clear meaning: it is the steady overtemperature of a far-away film region once all of its Joule heat leaves vertically into the diamond. With boundary conditions $\theta(0)=\theta_c$ (set jointly by Peltier pumping and the chip thermal network) and $\theta'(L_x)=0$ (adiabatic far end), the solution of Eq. \eqref{eq:fin} is:
\begin{equation}
\theta(x) = \theta_p + (\theta_c - \theta_p)\,\frac{\cosh[m(L_x - x)]}{\cosh(m L_x)}
\label{eq:fin-sol}
\end{equation}
Differentiate and evaluate the lateral energy flux at $x=0$. The Joule power recirculated to the cold electrode is:
\begin{equation}
Q_{ret} = k_{xy} L_z W\, m\,(\theta_p - \theta_c)\tanh(m L_x)
\label{eq:qret}
\end{equation}
where $W$ is the film width. The total Joule heat is $Q_J = q''' L_x L_z W = I^2 R_{TE}$. The recirculated fraction (recirculation share) is therefore:
\begin{equation}
\phi \equiv \frac{Q_{ret}}{Q_J}
= \frac{\tanh(m L_x)}{m L_x}\left(1 - \frac{\theta_c}{\theta_p}\right)
\label{eq:gamma}
\end{equation}
Equation \eqref{eq:gamma} is the central analytic result of this paper. Two factors, each with a clear physical meaning:
\begin{enumerate}
\item \textbf{Geometric suppression $\tanh(mL_x)/(mL_x)$}: when the film is much longer than the healing length ($mL_x \gg 1$), this factor $\approx l_p/L_x$. Before reaching the cold end, Joule heat needs to travel only about one healing length sideways, where the diamond captures it vertically. The returning share is set by the ratio of healing length to film length.
\item \textbf{Undercooling depth $(1-\theta_c/\theta_p)$}: a lateral temperature gradient toward the cold end exists only while Peltier pumping pulls $\theta_c$ below the Joule plateau $\theta_p$. The closer the cold end sits to the plateau temperature, the smaller the return. In the extreme of deep undercooling ($\theta_c \to 0$), the return reaches its ceiling.
\end{enumerate}
To match the compact-model convention of Eq. \eqref{eq:classic} (return term written as $\frac{\gamma}{2}I^2R$, with the classical vertical configuration at $\gamma = 1$), we define $\gamma \equiv 2\phi$. The recirculation factor is then bounded by:
\begin{equation}
\gamma_{\max} = \frac{2\tanh(m L_x)}{m L_x} \approx \frac{2 l_p}{L_x}
\quad (m L_x \gg 1)
\label{eq:gammamax}
\end{equation}
Furthermore, because the film--diamond path resistance is dominated by the film's own half-thickness ($R_{th,z}'' \approx L_z/2k_z$; TBR and spreading resistance are secondary terms), the healing length reduces to:
\begin{equation}
l_p = L_z\sqrt{\frac{k_{xy}}{2 k_z}}\;\xrightarrow{\;k_{xy}\approx k_z\;}\;\frac{L_z}{\sqrt{2}}
\label{eq:lp}
\end{equation}
This yields a \textbf{geometric decoupling rule} of direct design value:
\begin{equation}
\boxed{\;\gamma_{\max} \approx \sqrt{2}\,\frac{L_z}{L_x}\;}
\label{eq:designrule}
\end{equation}
That is: the recirculation factor is fixed by the film's \textbf{thickness-to-length ratio} alone, independent of thermal conductivities (material parameters still set the operating point through $\theta_p$ and $I_{opt}$). Making the film thinner and longer monotonically lowers the recirculation factor. This is the essential advantage of lying flat over standing upright: the vertical configuration locks $\phi$ at $1/2$ through internal resistance (Eq. \eqref{eq:pillar-split}), while the horizontal configuration lets a geometric ratio control $\phi$, with no floor.

\subsection{Order of Magnitude and Physical Picture of the Decoupling Rule}
\begin{figure*}[htbp]
\centering
\begin{tikzpicture}[font=\sffamily, scale=0.95, transform shape]
    \node[font=\large\bfseries] at (0, 5.2) {(a) Conventional vertical pillar (volume heat returns)};
    \filldraw[fill=orange!10, draw=orange!80, thick, rounded corners=1mm] (-2.5, 2.5) rectangle (2.5, 4);
    \node[orange!80!black, font=\bfseries] at (-1.5, 3.5) {GaN HEMT};
    \fill[red!60, opacity=0.4] (0, 2.8) ellipse (1.2cm and 0.4cm);
    \node[red!80!black, font=\bfseries] at (0, 2.8) {Extreme hotspot};
    \filldraw[fill=teal!15, draw=teal!80, thick] (-1, -0.5) rectangle (1, 2.5);
    \node[teal!60!black, font=\bfseries, rotate=90] at (-0.6, 1) {TE micro-bump (vertical)};
    \filldraw[fill=gray!10, draw=gray!80, thick, rounded corners=1mm] (-2.5, -2) rectangle (2.5, -0.5);
    \node[gray!80!black, font=\bfseries] at (0, -1.25) {Conventional PCB / heat sink};
    \draw[-{Stealth[length=3.5mm, width=2.5mm]}, blue!90, line width=2pt] (-0.2, 2.5) -- (-0.2, 1.5);
    \node[blue!90, anchor=east, font=\small\bfseries, align=right] at (-1.15, 2.0) {Peltier absorption};
    \fill[red!80] (0.3, 1) circle (2.5pt);
    \draw[-{Stealth[length=3.5mm, width=2.5mm]}, red!80, line width=2pt, densely dashed] (0.3, 1) -- (0.3, 2.4);
    \node[red!80, anchor=west, align=left, font=\small\bfseries] at (1.15, 1.7) {Joule heat return\\($\approx\frac{1}{2}i^2R$)};
    \draw[-{Stealth[length=3.5mm, width=2.5mm]}, red!80, line width=2pt] (0.3, 1) -- (0.3, -0.4);
    \node[red!80, anchor=west, align=left, font=\small\bfseries] at (1.15, 0.3) {Downward\\conduction};
    \begin{scope}[xshift=8.5cm]
        \node[font=\large\bfseries] at (0, 5.2) {(b) This work: horizontal decoupled (volume heat extracted vertically)};
        \filldraw[fill=orange!10, draw=orange!80, thick, rounded corners=1mm] (-3.5, 2.0) rectangle (3.5, 4);
        \node[orange!80!black, font=\bfseries] at (-2.5, 3.5) {GaN HEMT};
        \node[orange!80!black, font=\bfseries] at (0, 3.0) {Drain (electron exit)};
        \fill[red!60, opacity=0.4] (0, 2.3) ellipse (0.9cm and 0.25cm);
        \filldraw[fill=yellow!30, draw=yellow!80!orange, thick] (-3.5, 1.6) rectangle (3.5, 2.0);
        \node[orange!80!black, font=\small\bfseries] at (-2.6, 1.8) {DBC interface};
        \filldraw[fill=teal!15, draw=teal!80, thick] (-2, 0.8) rectangle (2, 1.6);
        \node[teal!60!black, font=\bfseries, align=center] at (-2.9, 1.2) {TE film\\(flat)};
        \filldraw[fill=cyan!10, draw=cyan!80, thick, rounded corners=1mm] (-3.5, -2) rectangle (3.5, 0.8);
        \node[cyan!80!black, font=\bfseries] at (0, -1.4) {Ultra-high-conductivity diamond ($>2000\,\mathrm{W/(m\cdot K)}$)};
        \draw[-{Stealth[length=3.5mm, width=2.5mm]}, blue!90, line width=2pt] (0, 2.0) -- (0, 1.1);
        \node[blue!90, anchor=west, align=left, font=\small\bfseries] at (0.3, 1.7) {Peltier interface absorption\\(localized at interface)};
        \foreach \x in {-1.2, -0.4, 0.4, 1.2} {
            \fill[red!80] (\x, 1.2) circle (2pt);
            \draw[-{Stealth[length=3mm, width=2mm]}, red!80, line width=1.5pt] (\x, 1.2) -- (\x, -0.1);
        }
        \node[red!80, font=\small\bfseries, align=center] at (0, -0.5) {Bulk Joule heat exits vertically into diamond\\(recirculation $\gamma \approx \sqrt{2}L_z/L_x \ll 1$)};
    \end{scope}
    \begin{scope}[yshift=-3.2cm, xshift=4.25cm]
        \filldraw[fill=white, draw=black!50, thick, rounded corners=2mm] (-4.6, -0.55) rectangle (4.6, 0.55);
        \node[font=\bfseries] at (-3.9, 0) {Legend:};
        \draw[-{Stealth[length=3mm, width=2mm]}, blue!90, line width=2pt] (-3.2, 0) -- (-2.2, 0);
        \node[right, font=\small\bfseries] at (-2.1, 0) {Peltier interface absorption path};
        \draw[-{Stealth[length=3mm, width=2mm]}, red!80, line width=2pt] (0.8, 0) -- (1.8, 0);
        \node[right, font=\small\bfseries] at (1.9, 0) {Joule volumetric heating path};
    \end{scope}
\end{tikzpicture}
\caption{Schematic comparison of internal heat-flow paths: a conventional vertical micro-bump thermoelectric module versus the horizontal spatially decoupled topology proposed here. The dashed red arrow shows Joule heat returning to the cold end in the classical structure --- a share locked near $1/2$ by the pillar's internal conduction resistance (Eq. \eqref{eq:pillar-split}). The right-hand topology uses geometric orthogonality and the low-resistance vertical path of diamond to press the recirculation factor down to $\gamma_{\max}\approx\sqrt{2}L_z/L_x$, set by the film's thickness-to-length ratio (Eq. \eqref{eq:designrule}).}
\label{fig:topology_comparison}
\end{figure*}
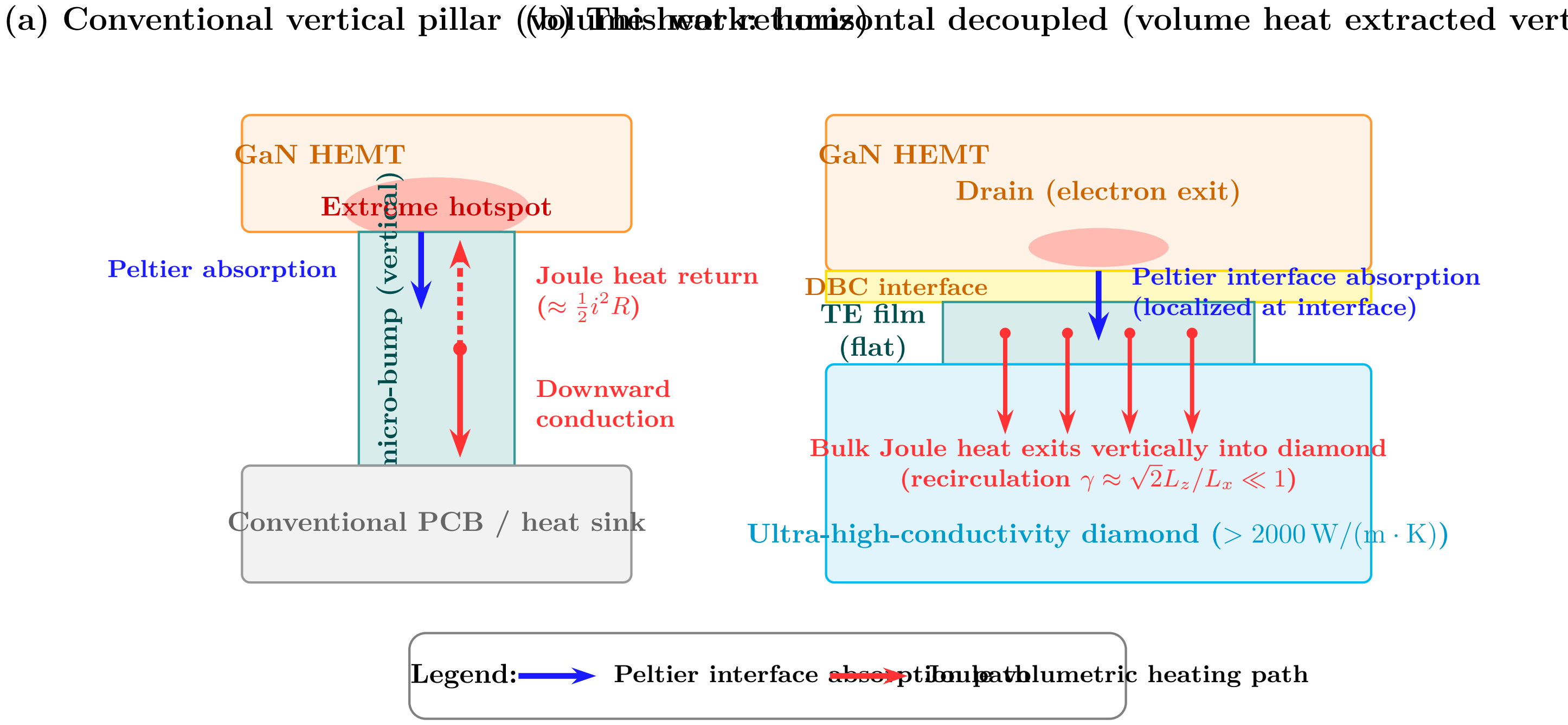

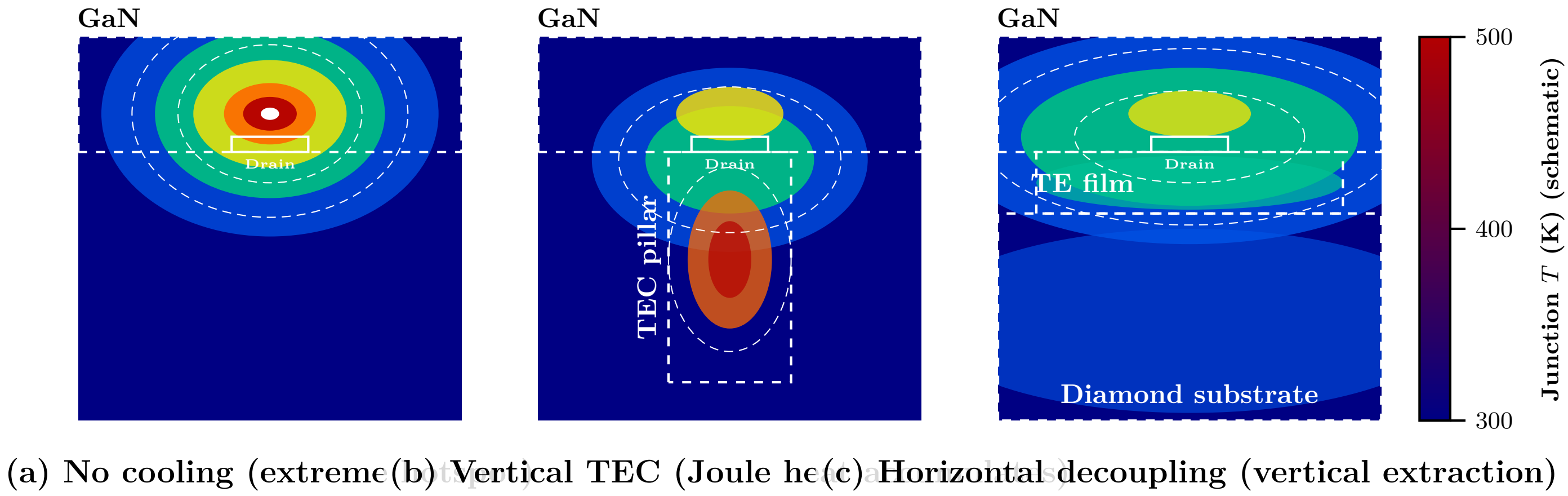
\begin{figure*}[htbp]
\centering
\begin{tikzpicture}[font=\sffamily, scale=0.9, transform shape]
    \definecolor{tblue}{RGB}{0,0,131}
    \definecolor{tcyan}{RGB}{0,105,255}
    \definecolor{tgreen}{RGB}{0,230,105}
    \definecolor{tyellow}{RGB}{255,230,0}
    \definecolor{torange}{RGB}{255,105,0}
    \definecolor{tred}{RGB}{180,0,0}
    \begin{scope}[xshift=0cm]
        \fill[tblue] (-2.5, -2.5) rectangle (2.5, 2.5);
        \begin{scope}
            \clip (-2.5, -2.5) rectangle (2.5, 2.5);
            \fill[tcyan, opacity=0.6] (0, 1.5) ellipse (2.2cm and 1.6cm);
            \fill[tgreen, opacity=0.7] (0, 1.5) ellipse (1.5cm and 1.1cm);
            \fill[tyellow, opacity=0.8] (0, 1.5) ellipse (1.0cm and 0.7cm);
            \fill[torange, opacity=0.9] (0, 1.5) ellipse (0.6cm and 0.4cm);
            \fill[tred, opacity=0.95] (0, 1.5) ellipse (0.35cm and 0.22cm);
            \fill[white] (0, 1.5) ellipse (0.12cm and 0.08cm);
            \draw[white, densely dashed, thin] (0, 1.5) ellipse (1.8cm and 1.35cm);
            \draw[white, densely dashed, thin] (0, 1.5) ellipse (1.2cm and 0.9cm);
        \end{scope}
        \draw[white, thick, dashed] (-2.5, 1.0) rectangle (2.5, 2.5);
        \node[black, font=\bfseries] at (-2.1, 2.75) {GaN};
        \draw[white, thick] (-0.5, 1.0) rectangle (0.5, 1.2);
        \node[white, font=\tiny\bfseries] at (0, 0.85) {Drain};
        \node[black, font=\large\bfseries, fill=white, inner sep=2pt,
              rounded corners=1mm, opacity=0.85, text opacity=1] at (0, -3.2) {(a) No cooling (extreme hotspot)};
    \end{scope}
    \begin{scope}[xshift=6cm]
        \fill[tblue] (-2.5, -2.5) rectangle (2.5, 2.5);
        \begin{scope}
            \clip (-2.5, -2.5) rectangle (2.5, 2.5);
            \fill[tcyan, opacity=0.6] (0, 0.9) ellipse (1.8cm and 1.2cm);
            \fill[tgreen, opacity=0.7] (0, 0.9) ellipse (1.1cm and 0.7cm);
            \fill[torange, opacity=0.75] (0, -0.4) ellipse (0.55cm and 0.9cm);
            \fill[tred, opacity=0.7] (0, -0.4) ellipse (0.28cm and 0.5cm);
            \fill[tyellow, opacity=0.8] (0, 1.5) ellipse (0.7cm and 0.35cm);
            \draw[white, densely dashed, thin] (0, 0.9) ellipse (1.45cm and 0.95cm);
            \draw[white, densely dashed, thin] (0, -0.4) ellipse (0.8cm and 1.2cm);
        \end{scope}
        \draw[white, thick, dashed] (-2.5, 1.0) rectangle (2.5, 2.5);
        \node[black, font=\bfseries] at (-2.1, 2.75) {GaN};
        \draw[white, thick] (-0.5, 1.0) rectangle (0.5, 1.2);
        \node[white, font=\tiny\bfseries] at (0, 0.85) {Drain};
        \draw[white, thick, dashed] (-0.8, -2.0) rectangle (0.8, 1.0);
        \node[white, font=\bfseries, rotate=90] at (-1.05, -0.5) {TEC pillar};
        \node[black, font=\large\bfseries, fill=white, inner sep=2pt,
              rounded corners=1mm, opacity=0.85, text opacity=1] at (0, -3.2) {(b) Vertical TEC (Joule heat accumulates)};
    \end{scope}
    \begin{scope}[xshift=12cm]
        \fill[tblue] (-2.5, -2.5) rectangle (2.5, 2.5);
        \begin{scope}
            \clip (-2.5, -2.5) rectangle (2.5, 2.5);
            \fill[tcyan, opacity=0.65] (0, 1.2) ellipse (3.2cm and 1.4cm);
            \fill[tgreen, opacity=0.7] (0, 1.2) ellipse (2.2cm and 0.9cm);
            \fill[tyellow, opacity=0.75] (0, 1.5) ellipse (0.8cm and 0.3cm);
            \fill[tgreen!70!tcyan, opacity=0.7] (0, 0.6) ellipse (2.0cm and 0.35cm);
            \fill[tcyan!80!tblue, opacity=0.6] (0, -1.2) ellipse (3.5cm and 1.2cm);
            \draw[white, densely dashed, thin] (0, 1.2) ellipse (2.7cm and 1.15cm);
            \draw[white, densely dashed, thin] (0, 1.2) ellipse (1.5cm and 0.6cm);
        \end{scope}
        \draw[white, thick, dashed] (-2.5, 1.0) rectangle (2.5, 2.5);
        \node[black, font=\bfseries] at (-2.1, 2.75) {GaN};
        \draw[white, thick] (-0.5, 1.0) rectangle (0.5, 1.2);
        \node[white, font=\tiny\bfseries] at (0, 0.85) {Drain};
        \draw[white, thick, dashed] (-2.0, 0.2) rectangle (2.0, 1.0);
        \node[white, font=\bfseries] at (-1.4, 0.6) {TE film};
        \draw[white, thick, dashed] (-2.5, -2.5) rectangle (2.5, 0.2);
        \node[white, font=\bfseries] at (0, -2.15) {Diamond substrate};
        \node[black, font=\large\bfseries, fill=white, inner sep=2pt,
              rounded corners=1mm, opacity=0.85, text opacity=1] at (0, -3.2) {(c) Horizontal decoupling (vertical extraction)};
    \end{scope}
    \begin{scope}[xshift=15.0cm, yshift=-2.5cm]
        \shade[bottom color=tblue, top color=tred] (0,0) rectangle (0.4, 5);
        \draw[black, thick] (0,0) rectangle (0.4, 5);
        \draw[thick] (0.4, 0) -- (0.6, 0) node[right, font=\small\bfseries] {$300$};
        \draw[thick] (0.4, 2.5) -- (0.6, 2.5) node[right, font=\small\bfseries] {$400$};
        \draw[thick] (0.4, 5) -- (0.6, 5) node[right, font=\small\bfseries] {$500$};
        \node[font=\small\bfseries, rotate=90, anchor=north] at (1.45, 2.5) {Junction $T$ (K) (schematic)};
    \end{scope}
\end{tikzpicture}
\caption{\textbf{Schematic} near-junction temperature fields for three architectures (pseudo-color is a qualitative mapping to illustrate heat-flow paths, not a quantitative simulation result). (a) Without cooling, phonon pile-up at the drain edge forms an extreme local hotspot. (b) A vertical TEC provides interface cooling, but volumetric Joule heat forms a secondary source inside the pillar and returns to the cold end in the proportion of Eq. \eqref{eq:pillar-split}. (c) The proposed topology extracts the film's volume heat through the low-resistance vertical path of diamond (a wide cool spread), so Peltier absorption acts mainly on the junction region. The decoupling is captured by the geometric ratio $\gamma_{\max}\approx\sqrt{2}L_z/L_x$.}
\label{fig:comsol_sim}
\end{figure*}

Take typical parameters (Table \ref{tab:params}): $L_z = 10\,\mathrm{\mu m}$, $L_x = 500\,\mathrm{\mu m}$, $k_{xy}=k_z=1.5\,\mathrm{W/mK}$, $R_{th,z}'' \approx 3.3\times10^{-6}\,\mathrm{m^2K/W}$. Then $l_p \approx 7\,\mathrm{\mu m}$, $mL_x \approx 71$, and
\begin{equation}
\gamma_{\max} \approx \frac{2 l_p}{L_x} \approx 0.028 \approx 0.03
\label{eq:gamma-num}
\end{equation}
Against the classical vertical $\gamma = 1$, a roughly $35$-fold reduction. Two caveats. First, $\gamma \to 0$ is not physically attainable: the healing length $l_p$ means that Joule heat generated within a band of width $\sim l_p$ around the cold electrode always has a part that returns. This residual coupling is thermodynamically irreducible. Second, once lattice-side leakage (Fourier back-flow from hot end to cold end) is included, the effective coefficient is $\gamma_{eff} \sim \gamma + 2l_p/L_x$ --- still controlled by the same geometric ratio. Figure \ref{fig:topology_comparison} compares the heat-flow paths of the two topologies; Figure \ref{fig:comsol_sim} sketches, in qualitative temperature fields, the difference among three architectures.

\subsection{Distributed-Multi-Physics Coupled Network Model}
\begin{figure}[htbp]
\centering
\begin{circuitikz}[font=\sffamily\footnotesize, scale=0.75, transform shape, thick]
    \draw (0,0) -- (8.8,0);
    \draw (8.8,0) node[rground]{};
    \node[below, font=\bfseries] at (7.4,-0.5) {$T_{amb}$ (system sink reference)};
    \draw (0.5,0) to[I, color=orange!90!black] (0.5,3.6);
    \node[orange!90!black, font=\bfseries, anchor=east, align=right] at (0.35,1.8) {Chip pulse\\source $P_{GaN}$};
    \draw (0.5,3.6) -- (7.6,3.6);
    \filldraw (2.4,3.6) circle (2.2pt);
    \node[above, font=\bfseries, align=center] at (2.4,3.8) {$T_{junc}$\\(electron-exit interface)};
    \draw (2.4,3.6) to[C] (2.4,0);
    \node[anchor=east, font=\bfseries] at (2.25,1.8) {$C_{th,j}$};
    \draw (4.2,3.6) to[cI, color=blue] (4.2,0);
    \node[blue, anchor=west, align=left, font=\bfseries] at (4.45,2.5) {$B_1=\Pi I_c$};
    \node[blue, anchor=west, align=left, font=\bfseries] at (4.45,1.6) {Interface Peltier\\transient absorption};
    \draw (4.2,3.6) to[R] (7.6,3.6);
    \node[font=\bfseries] at (5.9,3.25) {$R_{th,xy}$ (in-plane)};
    \filldraw (7.6,3.6) circle (2.2pt);
    \node[anchor=east, font=\bfseries] at (7.4,3.95) {$T_{film}$ (film body)};
    \draw (7.6,6.4) to[I, color=red] (7.6,3.6);
    \node[red, anchor=west, align=left, font=\bfseries] at (7.85,5.4) {$P_{Joule}$};
    \node[red, anchor=west, align=left, font=\bfseries] at (7.85,4.9) {$=\frac{\gamma}{2}i^2R$};
    \draw[cyan!80!black, line width=1.5pt] (7.6,3.6) to[R] (7.6,0);
    \node[cyan!80!black, anchor=west, align=left, font=\bfseries] at (7.85,2.2) {$R_{th,z}$};
    \node[cyan!80!black, anchor=west, align=left, font=\bfseries] at (7.85,1.6) {Diamond vertical\\low-impedance branch};
    \draw[-stealth, red, densely dashed, thick] (7.3,6.0) to[out=180, in=25] (4.0,4.4);
    \node[red, font=\bfseries, align=center] at (5.9,6.35) {Volume heat return suppressed ($\gamma\ll1$)};
\end{circuitikz}
\caption{Distributed-parameter Cauer thermal network model. In the electro-thermal analogy (current represents heat flow, voltage represents junction temperature): the controlled source $B_1$ models Peltier absorption at the interface. The high-conductivity diamond substrate creates a very low impedance branch $R_{th, z}$, so the volume heating ($P_{Joule}$, of which a fraction $\gamma/2$ returns to the cold side) drains vertically into the sink. The network topology suppresses Joule recirculation to the level $\gamma\ll1$.}
\label{fig:ltspice_circuit}
\end{figure}
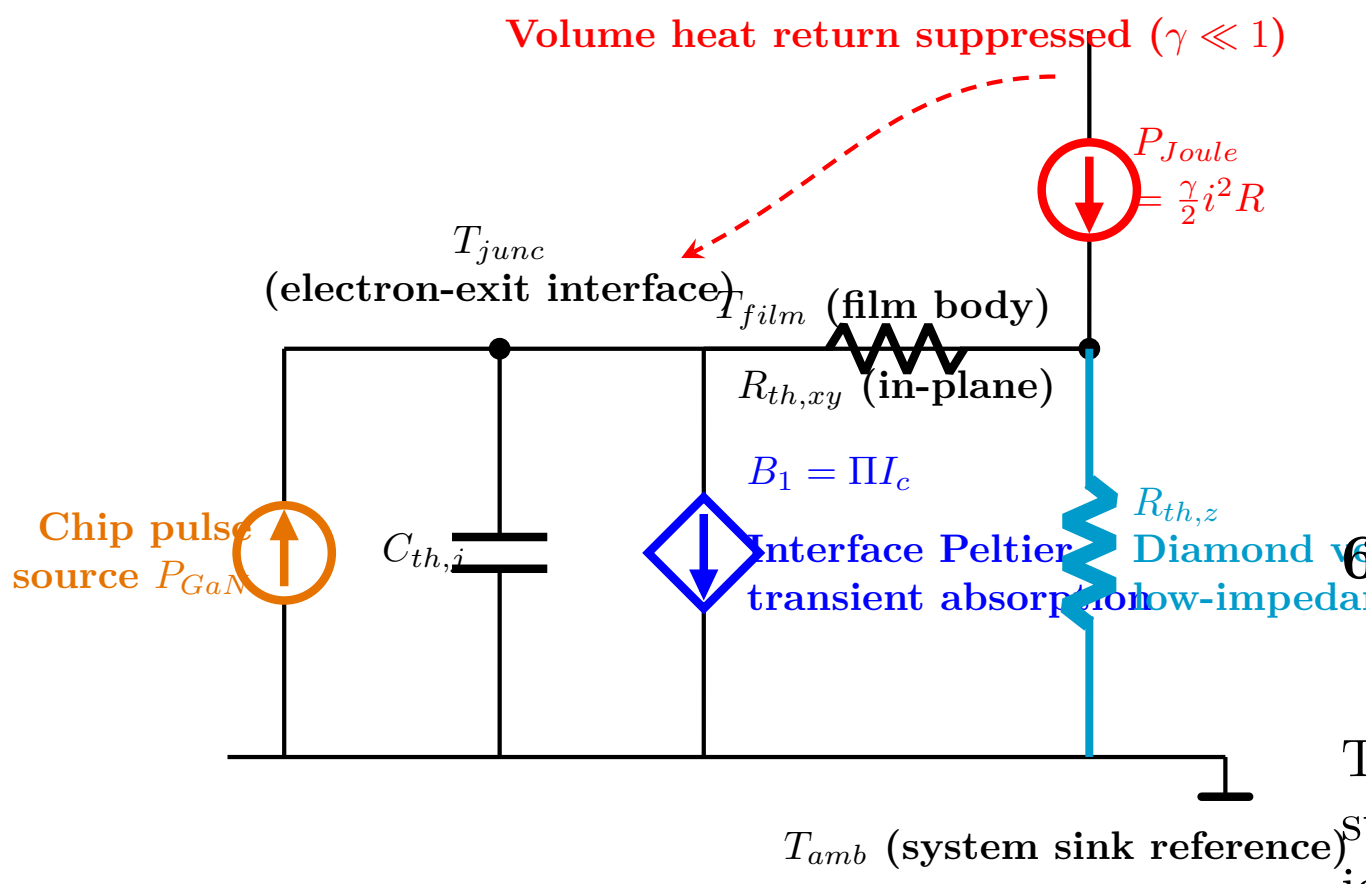

To describe transient conditions, we build a Cauer thermal network based on distributed parameters. The device is divided into micro-element cells; the node temperature vector $\bm{T}$ responds as:
\begin{equation}
\bm{C_{th}} \frac{d\bm{T}}{dt} + \bm{K_{th}} \bm{T} = \bm{P_{gen}} - \bm{P_{peltier}}
\label{eq:cauer}
\end{equation}
Here $\bm{K_{th}}$ is the three-dimensional conduction matrix. $\bm{P_{gen}}$ contains the intrinsic GaN heating and the film Joule heating (positive sources); $\bm{P_{peltier}}$ is the interface absorption vector (entering the equation with a negative sign by convention: absorption lowers the node enthalpy). In the nodal circuit, interface absorption is modeled as a controlled current sink, and the diamond path as a low-impedance branch to ground (see Figure \ref{fig:ltspice_circuit}). The network is the discrete counterpart of the continuum model in Section \ref{sec:gamma}: the $R_{th,z}$ branch is the vertical extraction path, the $R_{th,xy}$ branch is the lateral return path, and their ratio is exactly the decoupling mechanism captured by Eq. \eqref{eq:gamma}.

\section{Dynamic Operating Point, Stability, and Pulse Cooling}
Heat generation in high-power chips is strongly pulsed (radar RF, for instance). This section develops the time-domain and control-domain strategy. We first fix the scope of the temperature-sensitive parameters, then derive the generalized critical current $\lambda_m$ and the dynamic stability boundary. A time-constant analysis then sets the pulse cooling window, and a worked example closes the section.

\subsection{Temperature-Sensitive Parameters (TSP) and Their Scope}
The key parameters of a thermoelectric material or module are not constants. As the temperature difference $\Delta T$ between the ends fluctuates, the Seebeck coefficient and the resistance drift. Reference \cite{long2010optimization} gives a least-squares fit for a commercial multi-element TEC module (roughly hundreds of couples in series) over $\Delta T \in [0, 20]\,\mathrm{K}$ (\textbf{module-level} parameters, in V/K and $\Omega$):
\begin{equation}
Se(\Delta T) = 0.0518 - 2.14 \times 10^{-3} \cdot \Delta T \;\; [\mathrm{V/K}]
\label{eq:tsp1}
\end{equation}
\begin{equation}
R_0(\Delta T) = 5.24 + 2.32 \times 10^{-2} \cdot \Delta T + 7.09 \times 10^{-4} \cdot \Delta T^2 \;\; [\Omega]
\label{eq:tsp2}
\end{equation}
The boundaries of these fits matter. Eqs. \eqref{eq:tsp1}--\eqref{eq:tsp2} are \textbf{module-level}: the module's effective Seebeck coefficient sums over the series couples, so the values are far above the material-level $\sim 200\,\mathrm{\mu V/K}$. They hold only inside the fit window; extrapolation to large $\Delta T$ gives the unphysical result $\alpha \to 0$. The qualitative mechanism is general: as $\Delta T$ grows, the drift of $\alpha$ and the rise of resistance ($\mathrm{d}R_0/\mathrm{d}\Delta T > 0$) accelerate Joule heating, which destabilizes the system at high current. Section \ref{sec:stability} quantifies this. For an on-chip single-leg film device, one should use material-level $\alpha(T)$ and $\rho(T)$, fitted from superlattice film measurements. Here Eqs. \eqref{eq:tsp1}--\eqref{eq:tsp2} only illustrate the TSP feedback mechanism.

\subsection{Generalized Net Cooling Model and Derivation of the Critical Current $\lambda_m$}
\label{sec:lambda}
Generalize the classical model \eqref{eq:net-classic} to an arbitrary recirculation factor $\gamma$ (return term $\frac{\gamma}{2}I^2R$; $\gamma=1$ for the classical vertical case, $\gamma \approx 0.03$ for our horizontal film):
\begin{equation}
Q_{net}(I) = \alpha T_c I - \frac{\gamma}{2} R I^2 - K\Delta T
\label{eq:net-gamma}
\end{equation}
Net cooling requires $Q_{net} > 0$. Setting $Q_{net} = 0$ gives a quadratic in $I$, with roots:
\begin{equation}
I_{\pm} = \frac{\alpha T_c \pm \sqrt{(\alpha T_c)^2 - 2\gamma R K \Delta T}}{\gamma R}
\label{eq:roots}
\end{equation}
The system cools only between $I_-$ and $I_+$. Define the \textbf{critical current ceiling}:
\begin{equation}
\lambda_m \equiv I_{+} = \frac{\alpha T_c + \sqrt{(\alpha T_c)^2 - 2 \gamma R K \Delta T}}{\gamma R}
\label{eq:lambda}
\end{equation}
Equation \eqref{eq:lambda} generalizes the $\lambda_m$ of Ref. \cite{long2010optimization} to arbitrary recirculation factor (it reduces to the classical result at $\gamma=1$). Real roots require $\Delta T \le \Delta T_{max}(\gamma)$, where:
\begin{equation}
\Delta T_{max}(\gamma) = \frac{(\alpha T_c)^2}{2\gamma R K} = \frac{Z T_c^2}{2\gamma}
\label{eq:dtmax}
\end{equation}
Compare Eq. \eqref{eq:dtmax-classic}: \textbf{spatial decoupling scales both the maximum temperature difference and the net cooling ceiling by $1/\gamma$}. Likewise, the optimal current and maximum cooling become:
\begin{equation}
I_{opt} = \frac{\alpha T_c}{\gamma R},\qquad
Q_{max} = \frac{(\alpha T_c)^2}{2\gamma R} - K\Delta T
\label{eq:limits-gamma}
\end{equation}
The formal divergence as $\gamma \to 0$ needs a physical clarification. The divergence of $\lambda_m$ and $Q_{max}$ reflects the \textbf{lifting of the recirculation constraint}, not unlimited cooling. Three mechanisms cut off the real operating current: (i) the film's self-heating plateau $\theta_p = q''' L_z R_{th,z}'' \propto I^2$ grows with the square of current, limiting the effective temperature difference across the leg and the maintainable $ZT$; (ii) current density is bounded by electromigration and contact reliability; (iii) the undercooling depth is bounded by the Peltier--Fourier balance. The engineering meaning of $\gamma$ is therefore: \textbf{within a given current-density ceiling, it presses the erosion of net cooling by Joule heat down to a geometrically set minimum}. Infinite current is not the goal.

\subsection{Linear Stability and the Dynamic Runaway Boundary}
\label{sec:stability}
Beyond the static boundary \eqref{eq:lambda}, dynamic stability must also be examined. The energy balance for the junction temperature is $C_j\, dT_c/dt = P_{chip}(t) - Q_{net}(I, T_c, T_h)$. Perturb the equilibrium by $\delta T_c$ and linearize:
\begin{equation}
C_j \frac{d\,\delta T_c}{dt} = -\left(\frac{\partial Q_{net}}{\partial T_c}\right)\delta T_c + \cdots
\label{eq:perturb}
\end{equation}
The system is asymptotically stable exactly when the effective thermal conductance is positive:
\begin{equation}
\frac{\partial Q_{net}}{\partial T_c}
= I\left(\alpha + T_c\frac{\partial\alpha}{\partial T_c}\right)
-\frac{\gamma}{2}\frac{\partial R}{\partial T_c} I^2
+ K_{eff} \;>\; 0
\label{eq:stability}
\end{equation}
where $K_{eff}$ is the network's effective thermal conductance (including $K$ and its temperature-derivative terms). The physics in Eq. \eqref{eq:stability} is clear. The temperature dependence of the Peltier term ($\alpha + T_c\alpha'$) is usually self-stabilizing. The TSP-driven resistance slope $\partial R/\partial T_c > 0$ (typical of degenerate $\mathrm{Bi_2Te_3}$-family semiconductors) makes Joule heat feed back positively with temperature, and is the main destabilizer. Setting Eq. \eqref{eq:stability} to equality gives the dynamic boundary current:
\begin{equation}
I_{crit} = \frac{\left(\alpha + T_c\alpha'\right) + \sqrt{\left(\alpha + T_c\alpha'\right)^2 + 2\gamma R' K_{eff}}}{\gamma R'}
\label{eq:icrit}
\end{equation}
with $R' = \partial R/\partial T_c > 0$ and $\alpha' = \partial\alpha/\partial T_c$. Note that the dynamic boundary, like the static $\lambda_m$, scales as $1/\gamma$: \textbf{spatial decoupling relaxes the static cooling ceiling and the dynamic runaway boundary at the same time}. From the network viewpoint, Eq. \eqref{eq:stability} is the positive-definiteness condition of the Jacobian of the linearized Cauer network (Eq. \eqref{eq:cauer}): when $K_{eff}^{dyn} = \partial Q_{net}/\partial T_c$ changes sign, the matrix loses positive-definiteness and the system runs away. This gives a dynamical reading of the divergence criterion posed through convex programming in Ref. \cite{long2010optimization}.

\subsection{Characteristic Time Constants and the Pulse Cooling Window}
Radar and 5G base stations use RF pulses on the microsecond scale. Keeping the cooler on for long stretches accumulates needless Joule heat. We design the pulse strategy from the hierarchy of three characteristic time constants. Estimates based on Table \ref{tab:params} follow:
\begin{enumerate}
\item \textbf{Electron entropy transport response $\tau_{elec}$}: interface Peltier absorption follows the current on the electron relaxation time, picoseconds to nanoseconds. For engineering purposes, instantaneous.
\item \textbf{Film recirculation constant $\tau_{film}$}: Joule heat is \textbf{generated} instantly, but its temperature effect reaches the cold end only by lateral diffusion. With diffusivity $D_{th} = k/\,c_v \approx 1.5/1.2\times10^6 = 1.25\times10^{-6}\,\mathrm{m^2/s}$ and the healing length:
\begin{equation}
\tau_{film} \sim \frac{l_p^2}{D_{th}} \approx \frac{(7.1\times10^{-6})^2}{1.25\times10^{-6}} \approx 40\,\mathrm{\mu s}
\label{eq:tau-film}
\end{equation}
(The same value follows from the product of the film node heat capacity and the vertical resistance, $C''_{film} R_{th,z}'' = (c_v L_z)(L_z/2k_z)$; the two estimates agree.)
\item \textbf{Substrate dissipation constant $\tau_{sub}$}: the heat-capacity--resistance product of diamond and the package chain behind it, $\tau_{sub} \sim (c_v^{dia} t_{dia}) \cdot R_{TIM+coolant}'' \approx (1.8\times10^{6} \times 10^{-4}) \times 5\times10^{-4} \approx 0.09\,\mathrm{s}$.
\end{enumerate}
The hierarchy is:
\begin{equation}
\tau_{elec}\,(\sim\mathrm{ns}) \;\ll\; \tau_{film}\,(\sim 40\,\mathrm{\mu s}) \;\ll\; \tau_{sub}\,(\sim 0.1\,\mathrm{s})
\label{eq:tau}
\end{equation}
The pulse strategy follows directly. On detection of a current precursor (TSEP feedback), inject a cooling pulse of width $t_p \ll \tau_{film}$ ($1\sim10\,\mathrm{\mu s}$). During the pulse, interface Peltier absorption works at full power, while the film's Joule overtemperature builds only to $1 - e^{-t_p/\tau_{film}} \approx t_p/\tau_{film}$ ($3\%\sim25\%$) of its plateau. The instant high-power absorption at the interface cuts the peak junction temperature. After the pulse, the residual heat drains passively through the diamond on the $\tau_{sub}$ scale. This time-domain duty-cycle mechanism allows pulse amplitudes far above the steady optimum $I_{opt}$ (transient supercooling\cite{manno2015transient, yang2022transient}), while keeping the duty-cycle-averaged efficiency acceptable. One note: the supercooling depth is constrained by the undercooling factor $(1-\theta_c/\theta_p)$ in Eq. \eqref{eq:gamma}. Deeper undercooling raises the return share at the same time, so the optimal pulse amplitude is finite and should be tuned jointly with Eqs. \eqref{eq:lambda} and \eqref{eq:icrit}.

\subsection{Worked Example: Parameters and Quantitative Assessment}
\label{sec:example}
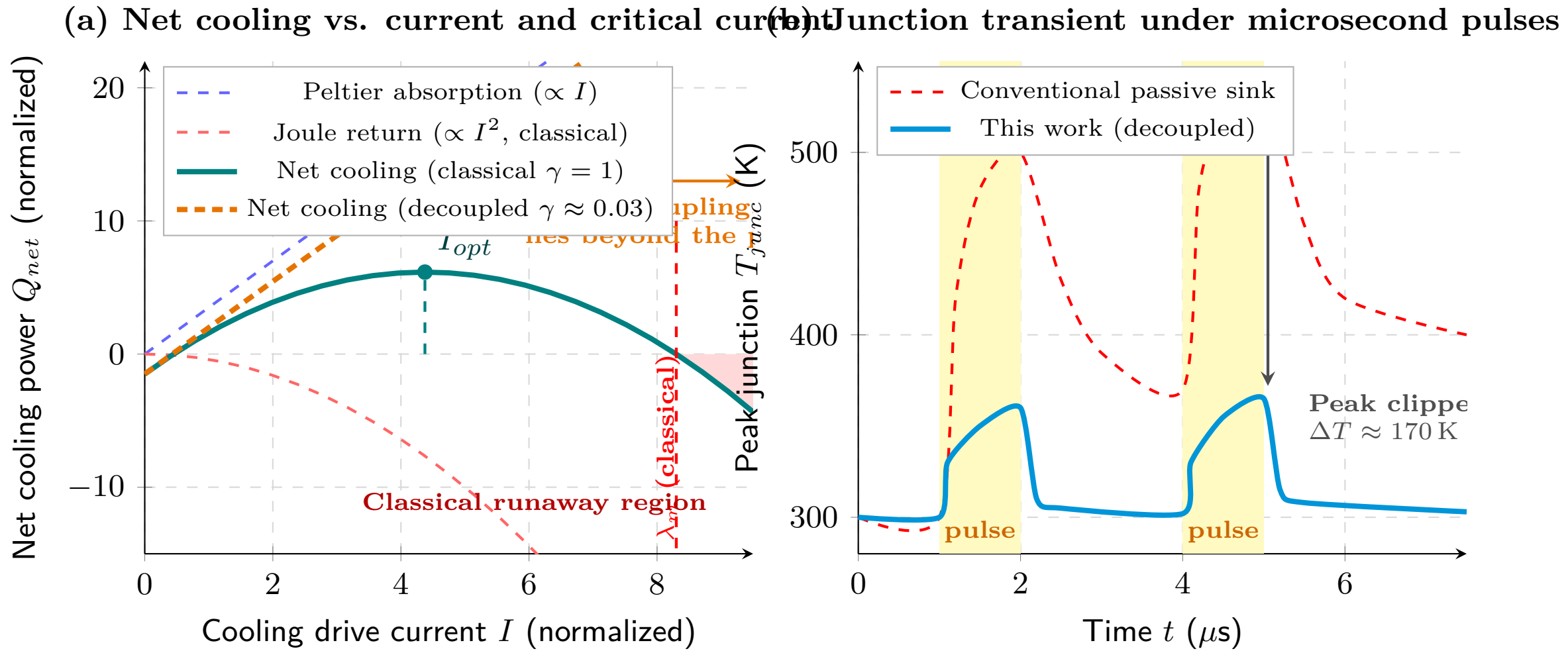
\begin{figure*}[!ht]
\centering
\begin{tikzpicture}[font=\sffamily\small]
    \begin{axis}[
        name=plotA,
        width=0.44\textwidth,
        height=6.5cm,
        title={(a) Net cooling vs. current and critical current},
        title style={font=\bfseries\small, yshift=-0.1cm},
        xlabel={Cooling drive current $I$ (normalized)},
        ylabel={Net cooling power $Q_{net}$ (normalized)},
        xmin=0, xmax=9.5,
        ymin=-15, ymax=22,
        clip=true,
        axis lines=left,
        grid=major,
        grid style={dashed, gray!30},
        legend style={at={(0.03,0.99)}, anchor=north west, font=\scriptsize, draw=gray!60, fill=white}
    ]
        \addplot [domain=0:9.5, blue!60, dashed, thick] {3.5*x};
        \addlegendentry{Peltier absorption ($\propto I$)}
        \addplot [domain=0:9.5, red!60, dashed, thick] {-0.4*x^2};
        \addlegendentry{Joule return ($\propto I^2$, classical)}
        \addplot [name path=net, domain=0:9.5, teal, line width=1.5pt] {3.5*x - 0.4*x^2 - 1.5};
        \addlegendentry{Net cooling (classical $\gamma=1$)}
        \addplot [name path=netd, domain=0:6.8, orange!90!black, line width=1.5pt, densely dashed] {3.5*x - 0.0112*x^2 - 1.5};
        \addlegendentry{Net cooling (decoupled $\gamma\approx0.03$)}
        \addplot [name path=zero, draw=none, domain=0:9.5] {0};
        \addplot[red!15] fill between[of=net and zero, soft clip={domain=8.3:9.5}];
        \coordinate (Opt) at (axis cs: 4.375, 6.156);
        \draw[dashed, teal, thick] (axis cs:4.375,0) -- (Opt);
        \filldraw[teal] (Opt) circle (2pt);
        \node[above right, font=\bfseries, teal!50!black] at (Opt) {$I_{opt}$};
        \draw[dashed, red, thick] (axis cs: 8.3, 10) -- (axis cs: 8.3, -15);
        \node[red, font=\scriptsize\bfseries, rotate=90, anchor=south] at (axis cs: 8.55, -7) {$\lambda_m$ (classical)};
        \node[red!70!black, font=\scriptsize\bfseries, align=left, anchor=south west] at (axis cs:3.2, -13) {Classical runaway region};
        \draw[-stealth, orange!90!black, thick] (axis cs: 6.9, 13) -- (axis cs: 9.3, 13);
        \node[orange!90!black, font=\scriptsize\bfseries, align=center, anchor=north] at (axis cs: 8.1, 12.7) {After decoupling, $\lambda_m$\\lies beyond the plot};
    \end{axis}
    \begin{axis}[
        at={(plotA.east)},
        anchor=west,
        xshift=1.05cm,
        width=0.44\textwidth,
        height=6.5cm,
        title={(b) Junction transient under microsecond pulses},
        title style={font=\bfseries\small, yshift=-0.1cm},
        xlabel={Time $t$ ($\mu$s)},
        ylabel={Peak junction $T_{junc}$ (K)},
        xmin=0, xmax=7.5,
        ymin=280, ymax=550,
        clip=true,
        axis lines=left,
        grid=major,
        grid style={dashed, gray!30},
        legend style={at={(0.03,0.99)}, anchor=north west, font=\scriptsize, draw=gray!60, fill=white}
    ]
        \fill[yellow!30] (axis cs:1,280) rectangle (axis cs:2,550);
        \fill[yellow!30] (axis cs:4,280) rectangle (axis cs:5,550);
        \node[orange!80!black, font=\scriptsize\bfseries, align=center] at (axis cs:1.5, 292) {pulse};
        \node[orange!80!black, font=\scriptsize\bfseries, align=center] at (axis cs:4.5, 292) {pulse};
        \addplot [red, thick, dashed, smooth, tension=0.5] coordinates {
            (0, 300) (1, 300)
            (1.2, 420) (1.5, 480) (2, 500)
            (2.5, 430) (3, 390) (4, 370)
            (4.2, 480) (4.5, 520) (5, 535)
            (5.5, 460) (6, 420) (7.5, 400)
        };
        \addlegendentry{Conventional passive sink}
        \addplot [cyan!80!blue, line width=1.5pt, smooth, tension=0.4] coordinates {
            (0, 300) (1, 300)
            (1.1, 330) (1.5, 350) (2, 360)
            (2.2, 310) (2.5, 305) (4, 302)
            (4.1, 330) (4.5, 355) (5, 365)
            (5.2, 315) (5.5, 308) (7.5, 303)
        };
        \addlegendentry{This work (decoupled)}
        \draw[-stealth, thick, black!70] (axis cs: 5.05, 528) -- (axis cs: 5.05, 372);
        \node[black!70, font=\scriptsize\bfseries, align=left, anchor=west] at (axis cs: 5.4, 355) {Peak clipped\\$\Delta T\approx170\,\mathrm{K}$};
    \end{axis}
\end{tikzpicture}
\caption{Critical-current search and transient junction response under pulse cooling (compact-model predictions from Eqs. \eqref{eq:net-gamma}--\eqref{eq:lambda}. The vertical axis in (a) is normalized to compare the two configurations; the ratio of curve coefficients matches the $\gamma$ values in Table \ref{tab:params}. Not experimental data.) (a) The contest between the interface term ($\propto I$) and the volume return term ($\propto I^2$) gives the classical configuration an optimum $I_{opt}$ and a divergence boundary $\lambda_m$ (the red-shaded part: negative net cooling). With $\gamma\approx0.03$ the return term drops by more than an order of magnitude, $\lambda_m$ moves beyond the plotted range, and net cooling at the same current rises sharply. (b) Under microsecond high-power pulses, a conventional passive sink shows baseline drift and heat accumulation. The proposed topology combines instant interface pumping with fast spreading in diamond, and the model predicts a strong suppression of peak junction temperature (the shown clipping is a model prediction and depends on pulse tuning).}
\label{fig:pulse_cooling}
\end{figure*}

Table \ref{tab:params} lists the geometry, materials, and derived parameters used in the example. The comparison holds the resistance fixed (same $\alpha$, $R$, $K$; only the recirculation factor differs). Figure \ref{fig:pulse_cooling}(a) shows the net cooling versus current for the classical and the decoupled configurations.

\begin{table*}[htbp]
\centering
\caption{Worked-example parameters (N-type $\mathrm{Bi_2Te_3}$ superlattice film, electrically isolated reference design)}
\label{tab:params}
\begin{tabular}{llll}
\toprule
Symbol & Meaning & Value & Source \\
\midrule
$\alpha$ & Seebeck coefficient (magnitude) & $2.4\times10^{-4}\,\mathrm{V/K}$ & Order of magnitude from Ref. \cite{bulman2016superlattice} \\
$\sigma$ & Electrical conductivity & $1.0\times10^{5}\,\mathrm{S/m}$ & Same as above \\
$k_{TE}$ & Film thermal conductivity & $1.5\,\mathrm{W/(m\cdot K)}$ & Low $\kappa$ of superlattice \\
$L_z$ & Film thickness & $10\,\mathrm{\mu m}$ & Design value \\
$L_x$ & Lateral length, cold to far electrode & $500\,\mathrm{\mu m}$ & Design value \\
$W$ & Film width & $2\,\mathrm{mm}$ & Design value \\
$R_{TE}$ & Film resistance $\rho L_x/(L_z W)$ & $0.25\,\mathrm{\Omega}$ & Calculated \\
$R_{th,z}''$ & Film--diamond areal thermal resistance & $\approx3.3\times10^{-6}\,\mathrm{m^2K/W}$ & Calculated (TBR $\sim10^{-8}$ \cite{cheng2020interfacial}) \\
$K$ & Film conductance $k_{xy}L_zW/L_x$ & $6.0\times10^{-5}\,\mathrm{W/K}$ & Calculated \\
$l_p$ & Lateral thermal healing length & $\approx7\,\mathrm{\mu m}$ & Eq. \eqref{eq:lp} \\
$\gamma_{\max}$ & Recirculation ceiling $2l_p/L_x$ & $\approx0.028$ & Eq. \eqref{eq:gammamax} \\
$T_c$ & Cold-side operating temperature & $350\,\mathrm{K}$ & Operating condition \\
$\Delta T_{leg}$ & Temperature drop across the leg & $\approx25\,\mathrm{K}$ & Calculated ($\approx \theta_p$) \\
$Q_0$ & Design cooling load & $0.35\,\mathrm{W}$ & $5\,\mathrm{kW/cm^2}\times$ hotspot area \\
\bottomrule
\end{tabular}
\end{table*}

The quantitative assessment based on Table \ref{tab:params}:
\begin{enumerate}
\item \textbf{Maximum cooling}: classical ($\gamma=1$), $Q_{max} = (\alpha T_c)^2/(2R) - K\Delta T_{leg} \approx 14.1 - 1.5 = 12.6\,\mathrm{mW}$. Decoupled ($\gamma=0.028$), $Q_{max} \approx 502\,\mathrm{mW}$. The same film, merely re-oriented, gains about $40$ times in cooling capacity and covers the $0.35\,\mathrm{W}$ design load; the classical configuration cannot.
\item \textbf{Critical current}: by Eq. \eqref{eq:lambda}, $\lambda_m^{classic} \approx 0.65\,\mathrm{A}$ and $\lambda_m^{decoupled} \approx 24\,\mathrm{A}$, an extrapolation of about $37$ times.
\item \textbf{Operating point}: solving $Q_{net}(I) = Q_0$ gives $I \approx 5.4\,\mathrm{A}$, below $I_{opt} = \alpha T_c/(\gamma R) \approx 12\,\mathrm{A}$, i.e., on the rising branch of net cooling. The current density is $J \approx 2.7\times10^{4}\,\mathrm{A/cm^2}$, within the reported working range of thin-film thermoelectric devices. The film's Joule plateau overtemperature is then $\theta_p = \rho J^2 L_z R_{th,z}'' \approx 24\,\mathrm{K}$, consistent with $\Delta T_{leg}$; the material $ZT$ is maintained.
\item \textbf{An honest look at efficiency}: at continuous duty, $\mathrm{COP} = Q_0/(\alpha I\Delta T_{leg} + I^2R) \approx 0.35/7.3 \approx 0.05$. This is low. It is the intrinsic price of pumping at extreme heat flux: the Joule loss no longer returns to the cold end, but it still counts against electrical power. The value of the architecture lies in the \textbf{reachability of cooling capacity} and in \textbf{duty-cycle efficiency under pulses}, not in a high continuous COP.
\end{enumerate}

\section{Material Selection and Interface Phonon Engineering}
Theory lands only if materials and interface processes support it. This section discusses the physical and chemical prerequisites of the configuration.

\subsection{Matching the Thermoelectric Figure of Merit ($ZT$) Across Hetero-Materials}
The topology asks two things of the materials: a large Seebeck difference, to strengthen the pumping interface; and low resistance, to weaken the heat source. The standard metric is the dimensionless figure of merit $ZT$:
\begin{equation}
ZT = \frac{\alpha^2 \sigma}{\kappa} T
\end{equation}
For near-room-temperature operation ($300\,\mathrm{K}\sim 400\,\mathrm{K}$), superlattice alloys based on bismuth--tellurium ($\mathrm{Bi_xSb_{1-x}Te_3}$) are the natural choice. In modern molecular beam epitaxy (MBE) superlattices, phonons scatter diffusely at the many layer interfaces, which suppresses the lattice conductivity $\kappa_l$; electrons keep a high $\sigma$ through level tunneling. The combination breaks the Wiedemann--Franz constraint that binds bulk materials\cite{bulman2016superlattice}. Note that Eq. \eqref{eq:designrule} shows the recirculation factor does not depend on thermal conductivity. Material work therefore targets $\alpha$ and $\sigma$ (the power factor $\alpha^2\sigma$), while geometry ($L_z/L_x$) carries the recirculation suppression on its own. In this framework the material and geometric degrees of freedom are decoupled, which simplifies co-design.

\subsection{DBC Processing and Suppression of Thermal Boundary Resistance (TBR)}
At the junction of dissimilar materials, the phonon density of states does not match, and a thermal boundary resistance (TBR) appears. Per the acoustic mismatch model (AMM), nanoscale air gaps or soft van der Waals contact --- between the film and the drain metal, or between the film and the diamond below --- would create interface resistance large enough to block the transient heat pumping.

The architecture therefore calls for metallurgical-grade processes: direct bonded copper (DBC) and plasma activated bonding. Sputter a Ti/Pt/Au interlayer on the diamond and grow the alloy in situ, removing physical gaps as far as possible. TBR cannot reach zero --- its floor is set by phonon mismatch, at the $\sim10^{-8}\,\mathrm{m^2K/W}$ scale. But Eq. \eqref{eq:lp} shows when TBR dominates: only if $R_{th,z}'' \ll L_z/2k_z$. For our film geometry ($L_z/2k_z \approx 3.3\times10^{-6}\,\mathrm{m^2K/W}$), TBR at today's process level is not the bottleneck for decoupling performance. The bar for engineering implementation is lowered\cite{khanna2017thin}.

Embedding a thermoelectric structure near the electron exit of an RF device still faces two things: phonon mismatch at hetero-interfaces, and parasitic capacitance perturbing the high-frequency signal (see the quantitative assessment in Section \ref{sec:interface}). Future implementations can rely on more advanced room-temperature plasma bonding to reduce lattice defects, and on conformal electrode design or field-plate reuse, which controls the parasitics while preserving the intrinsic microwave performance of the GaN device.

\section{Conclusions and Outlook}
Moving core heat out of the core fast enough, and holding the core temperature line, is a shared problem for large-scale integrated circuits and high power-density devices. Against the extreme local hotspots of $5\,\mathrm{kW/cm^2}$ inside modern GaN HEMTs, macroscopic passive cooling based on a conduction chain has reached its physical limit. This paper starts from a first principle --- the electron carries both charge and entropy --- and proposes and systematically develops a horizontal near-junction active cooling architecture based on spatial decoupling.

The analytic model and the worked example use the extreme case of a GaN HEMT, but the underlying mechanism --- separating the interfacial Peltier source from the volumetric Joule source in space --- is a general rule of nonequilibrium thermodynamics. The topology does not alter the intrinsic semiconductor structure of the device. In principle it transfers to ultra-high-compute silicon multi-chip modules (MCM), core hotspot suppression in CPUs and GPUs, SiC power electronics, and photonic integrated circuits (PIC). Any chip that must move core heat out of its core can benefit from the spatial decoupling principle.

The core conclusions and theoretical contributions are:
\begin{enumerate}
\item \textbf{Rigor at the mechanism level}: within the Onsager transport framework, a strict definition of ``electron entropy transport'' is given --- the entropy flow $\bm{J}_s^e = \alpha\bm{J}_e$, with the Seebeck coefficient as entropy transported per unit charge. The energy equation with current, $\nabla\cdot(k\nabla T) + \rho J^2 - T\bm{J}\cdot\nabla\alpha = 0$, proves the origin of the spatial asymmetry: interface localization of the Peltier effect and volume distribution of Joule heating. The basis of the Kelvin relation $\Pi = \alpha T$ is also completed.
\item \textbf{An analytic recirculation factor and a geometric design rule}: the exact split formula for the vertical pillar proves that the classical return share is locked at $1/2$ by the internal conduction resistance and cannot be removed by interface engineering. The lateral fin solution for the flat film gives the recirculation factor $\gamma = \frac{\tanh(mL_x)}{mL_x}\left(1-\frac{\theta_c}{\theta_p}\right)$ and the geometric rule $\gamma_{\max} \approx \sqrt{2}L_z/L_x$, turning recirculation suppression into a precisely designable geometric parameter problem. In the equal-resistance comparison, maximum cooling rises from about $13\,\mathrm{mW}$ to about $500\,\mathrm{mW}$ (roughly $40$ times).
\item \textbf{Safety boundaries and dynamic stability}: the generalized critical current $\lambda_m$ and maximum temperature difference $\Delta T_{max}(\gamma) = ZT_c^2/(2\gamma)$ are derived; the physical cutoffs behind the formal divergence at $\gamma\to0$ are clarified (self-heating plateau $\theta_p \propto I^2$, current density, electromigration). Linearizing the Cauer network gives the dynamic runaway boundary $I_{crit}$, and both static and dynamic boundaries scale as $1/\gamma$. The time-constant hierarchy ($\mathrm{ns} \ll 40\,\mathrm{\mu s} \ll 0.1\,\mathrm{s}$) fixes the working window for microsecond pulse cooling.
\end{enumerate}

The model also contains known simplifications, which point to future work: (i) the temperature dependence of TBR and material parameters, and the coupling of thermal stress and reliability (electromigration, thermal-cycling fatigue), are not included; (ii) the interface parasitics (overlap capacitance, interconnect inductance) need full-wave electromagnetic simulation and RF measurement for verification; (iii) the lateral fin model is one-dimensional --- two-dimensional effects near the cold electrode and the spreading resistance in diamond need a three-dimensional finite-element check; (iv) all quantitative conclusions here are analytic and compact-model predictions, pending experimental verification on superlattice film samples. These form the focus of the next stage of experiment and refined multi-physics modeling.

In summary, the self-cooled transistor concept offers an analytic design path --- strict in mechanism, computable in parameters, bounded in margins --- for the micro-scale thermal bottleneck from power transistors to large-scale integrated circuits, and lays a physical foundation for the next generation of chip-level intrinsic thermal architectures.

\end{document}